\documentclass[aps,pra,epsfigure,twocolumn,superscriptaddress]{revtex4-1}
\usepackage[colorlinks=true,linkcolor=blue,urlcolor=blue,citecolor=blue,pdfusetitle]{hyperref}
\usepackage[utf8]{inputenc}
\usepackage[english]{babel}
\usepackage{amsmath}
\usepackage[caption = false]{subfig}
\usepackage{graphicx,epstopdf}
\usepackage{blindtext}
\usepackage[table,xcdraw]{xcolor}
\usepackage{lipsum}
\usepackage{amsfonts}
\usepackage{bbm}
\usepackage{amssymb}
\usepackage{enumerate}
\usepackage{color}
\usepackage{latexsym}
\usepackage{physics}
\usepackage{times,txfonts}
\newtheorem{theorem}{Theorem}

\newcommand{\Dcal}{\mathcal{D}}

\newcommand{\Fcal}{\mathcal{F}}

\newcommand{\Ical}{\mathcal{I}}

\newcommand{\Ocal}{\mathcal{O}}

\newcommand{\Scal}{\mathcal{S}}

\newcommand{\Ucal}{\mathcal{U}}

\newcommand{\1}{\mathbbm{1}}

\newcommand{\Rmath}{\mathbbm{R}}

\newcommand{\interpro}[2]{\langle #1 | #2 \rangle}

\newcommand{\SubFig}[2]{\ref{#1}{\color{blue}#2}}

\definecolor{bluePoli}{cmyk}{0.4,0.1,0,0.4}
\definecolor{blueGreen}{RGB}{0, 102, 102}
\definecolor{brickred}{rgb}{0.8, 0.25, 0.33}
\definecolor{darkred}{RGB}{204, 0, 0}
\definecolor{darkgreen}{RGB}{0, 102, 50}
\definecolor{darkblue}{RGB}{0, 76, 153}
\definecolor{mygold}{RGB}{139, 135, 32}

\newcommand{\revisionRefB}[1]{{\color{black}#1}}

\newcommand{\CSIC}{Instituto de Física Fundamental (IFF), Consejo Superior de Investigaciones Científicas (CSIC), Calle Serrano 113b, 28006 Madrid, Spain.}

\usepackage{orcidlink}

\begin{document}

\title{Validity condition for high-fidelity Digitized Quantum Annealing}

\author{Alan C. Santos~\orcidlink{0000-0002-6989-7958}}
\email{ac\_santos@iff.csic.es}
\affiliation{\CSIC}


\begin{abstract}
	Digitizing an adiabatic evolution is a strategy able to combine the good performance of gate-based quantum processors with the advantages of adiabatic algorithms, providing then a hybrid model for efficient quantum information processing. In this work we develop validity conditions for high fidelity digital adiabatic tasks. To this end, we assume a digitizing process based on the Suzuki-Trotter decomposition, which allows us to introduce a \textit{Digitized Adiabatic Theorem}. As consequence of this theorem, we show that the performance of such a hybrid model is limited by the fundamental constraints on the adiabatic theorem validity, even in ideal quantum processors. We argue how our approach predicts the existence of intrinsic non-adiabatic errors reported by R. Barends \textit{et al.}, Nature \textbf{534}, 222 (2016) through an empirical study of digital annealing. In addition, our approach allows us to explain the existence of a scaling of the number of Suzuki-Trotter blocks for the optimal digital circuit with respect to the optimal adiabatic total evolution time, as reported by G. B. Mbeng \textit{et al.}, Phys. Rev. B \textbf{100}, 224201 (2019) through robust numerical analysis of digital annealing. We illustrate our results through two examples of digitized adiabatic algorithms, namely, the two-qubits exact-cover problem and the three-qubits adiabatic factorization of the number 21.
\end{abstract}

\maketitle

\section{Introduction}

Analog quantum computers and simulators are devices that use tunable and continuous varying parameters, like fields and interactions, of a programmable quantum system to process information, as idealized by Prof. Richard Feynman in his seminal paper~\cite{Feynman:82}. In this context, adiabatic evolution~\cite{Kato:50,Born:28} is a strategy widely used for optimization and satisfiability problems through quantum annealing processors~\cite{Tadashi:98,Farhi:00,Farhi:01}, like the analog D-Wave Systems' processors~\cite{Johnson:11,Pudenz:14,King:22}, and trapped atoms based quantum simulators~\cite{Feng:23,Chen:23}. On the other hand, the gate-based model of quantum computation offers a number of different benefits thanks to the existence of a finite number of gates required for universal quantum computation~\cite{Barenco:95}, in addition to taking advantage of quantum error correction codes~\cite{Peres:85,Shor:95}, among others~\cite{Nielsen:Book}. Different from analog version for computation, the gate-based model exploits the application of a discrete amount of gates in a circuit-based approach, and we call it \textit{digital} model for quantum information processing. 

Because each of the models aforementioned have its own set of advantages and disadvantages, a scheme for quantum computation based on the digitization of adiabatic evolutions seems to be a good route for a new hybrid and robust proposal 
for quantum control and information processing~\cite{Steffen:03,Shapiro:07}. 
In this regard, Barends \textit{et al}~\cite{Barends:16} introduced the notion of \textit{digitized adiabatic quantum computing} (DAQC), a hybrid model able to take advantage of the benefits of digital computation to suppress some errors related to operations of analog computers. In such a work, they experimentally investigated the performance of DAQC for a particular evolution of a spin chain (with up to 9 spins), where the performance of the model was done by simulating the time-continuous adiabatic evolution of the system though a length-fixed quantum circuit. Motivated by this experiment, Mbeng \textit{et al}~\cite{Mbeng:19} provided an interesting analysis (benchmark) to estimate the minimum required number of blocks, $N_{\mathrm{bl}}$, to digitize an adiabatic evolution with high fidelity. 

The Refs.~\cite{Barends:16,Mbeng:19} focused their analysis to different spin models, but their conclusion are mutually consistent. Of particular interest of this work, both works report that the fidelity (performance) of the analog adiabatic evolution and its digitized version are identically unsatisfactory for small total evolution time $\tau$ (fast dynamics). In addition, even for slow dynamics (large $\tau$), the analog adiabatic evolution has a performance better than its digitized counterpart. In this regard, in this work we propose a theoretical description for DAQC able to explain these results reported in the works~\cite{Barends:16,Mbeng:19}, in addition to predict other important results in this context. To this end, we introduce the \textit{Suzuki-Trotter digitized adiabatic theorem} (ST-DAT), a sufficient condition for high fidelity DAQC based on the Suzuki-Trotter expansion. We exemplify the consequences of the ST-DAT in two different algorithms based on the Ising Hamiltonian. 


\section{Suzuki-Trotter Digitized Adiabatic Theorem}

In this section we introduce a sufficient condition for the approximate simulation of analog adiabatic quantum tasks in digital quantum computers. But, first of all, let us recall the quantum adiabatic theorem and its validation rules~\cite{Sarandy:04,Amin:09}. Consider the dynamics of a quantum system ruled by the Schrödinger equation
\begin{align}
	i\hbar \ket*{\dot{\psi}(t)} = \hat{H}(t)\ket{\psi(t)}, \label{Eq:Schrodinger}
\end{align}
in the interval $t \in [0,\tau]$, where the system is governed by a time-dependent Hamiltonian $H(t)$ satisfying the instantaneous eigenvalue equation $\hat{H}(t)\ket{E_{n}(t)} = E_{n}(t)\ket{E_{n}(t)}$. By starting the system in an arbitrary initial state $\ket{\psi(0)} = \sum_{n}^{\Dcal} c_{n} \ket{E_{n}(0)}$ ($\Dcal$ the dimension of the Hilbert space of the system), one says that a quantum system will evolves under the adiabatic trajectory, $\ket{\psi_{\mathrm{ad}}(t)} = \sum\nolimits_{n=1}^{\Dcal} c_{n} e^{-\frac{i}{\hbar} \int_{t_{0}}^{t} E_{n}(t) + i\interpro{\dot{E}_{n}(t)}{E_{n}(t)} } \ket{E_{n}(t)}$, when the total evolution time of the dynamics obeys the following rigorous inverse gap squared criteria (by using the definition of the normalized time $s = t/\tau$)~\cite{Sarandy:04,Amin:09,Jansen:07,Cao:13,Lidar:09,Messiah:Book,Berry:84,Berry:09}
\begin{align}
 \tau \geq \tau_{\mathrm{ad}} = \max_{0\leq s \leq 1} \left[\hbar \norm{d_{s}H(s)}/\Delta^2(s)\right] , \label{Eq:AdTime}
\end{align}
where $\Delta(s)$ is the minimum gap energy between all energy states of the Hamiltonian. We use $\norm{\bullet} = \sqrt{\mathrm{tr}(\bullet^{\dagger}\bullet)}$ to denote the Hilbert-Schmidt norm, and $d_{s}$ is the derivative with respect to the time $s$. 
	It is worth stating that the above condition is not the most general and rigorous condition for adiabaticity. In fact, as stated by in Ref.~\cite{Jansen:07}, we can derive conditions based on higher orders of $\norm{d_{s}H(s)}$ and $\Delta(s)$. However, the above condition can be used to estimate $\tau_{\mathrm{ad}}$ for a large class of Hamiltonians~\cite{Rezakhani:09}.

We refer to $\tau_{\mathrm{ad}}$ as the \textit{adiabatic speed threshold}. It is timely to say that the above condition is valid for an arbitrary input state $\ket{\psi(0)}$, since a maximization over the entire energy spectrum of the Hamiltonian~\cite{Amin:09} is implicitly included in this condition. For the class of problems to be consider here the above condition is \textit{sufficient}, but \textit{it is not necessary}. In other words, the adiabatic behavior can be achieved without satisfying this condition; however, if it is satisfied, we immediately conclude that the system dynamics will be (approximately) the adiabatic trajectory $\ket{\psi_{\mathrm{ad}}(t)}$.

Without loss of generality, we will assume in this manuscript the class of generic adiabatic Hamiltonians of the form
\begin{align}
	\hat{H}(s) = f(s) \hat{H}_{\mathrm{ini}} + g(s) \hat{H}_{\mathrm{fin}} , \label{Eq:Hamiltonian}
\end{align}
for arbitrary initial $\hat{H}_{\mathrm{ini}}$ and final Hamiltonians $\hat{H}_{\mathrm{fin}}$, such that the functions $f,g: \Rmath \rightarrow \Rmath$ satisfy $f(0)=g(1)=1$ and $f(1)=g(0)=0$. In this way, the adiabatic evolution operator can be written as $U_{\mathrm{ad}}(s) \approx U(s)\vert_{\tau \gg \tau_{\mathrm{ad}}}$, where $U(s)\vert_{\tau \gg \tau_{\mathrm{ad}}}$ denotes the solution of the Eq.~\eqref{Eq:Schrodinger} for a sufficiently slow-varying time-dependent Hamiltonian (when $\tau \gg \tau_{\mathrm{ad}}$). Our strategy to digitize the dynamics considers a sequence of time intervals $\delta s_{n} = s_{n+1}-s_{n}$ such that the evolution is separated in $M$ sub-intervals $\Scal_{n} = [s_{n},s_{n+1}]$. In this case, according to digitized quantum computation method we write the adiabatic evolution operator as (with $s_{0}=0$)
\begin{align}
	\hat{U}_{\mathrm{ad}}(s) = \lim_{M\rightarrow \infty} \left[\hat{U}_{\mathrm{d}}(s_{M};s_{M-1}) \cdots \hat{U}_{\mathrm{d}}(s_{2};s_{1})\hat{U}_{\mathrm{d}}(s_{1};s_{0})\right] , \label{Eq:U_dig}
\end{align}
with $\hat{U}_{\mathrm{d}}(s_{n+1};s_{n})$ the evolution operator for the $n$-th digitized block. However, in order to avoid the limit $M\rightarrow \infty$, we want to estimate the finite number $M_{\mathrm{ad}}$ which allows us to write
\begin{align}
	\hat{U}_{\mathrm{ad}}(s) \approx \hat{U}_{\mathrm{d}}(s_{M_{\mathrm{ad}}};s_{M_{\mathrm{ad}}-1}) \cdots \hat{U}_{\mathrm{d}}(s_{2};s_{1})\hat{U}_{\mathrm{d}}(s_{1};s_{0}) .
\end{align}

To this end, given the nature of the evolution operator as a time-ordering integration~\cite{Sakurai:Book}, we propose to use a convenient way to compute each block $\hat{U}_{\mathrm{d}}(s_{n+1};s_{n})$ through what we call \textit{Riemann-like digitization}. That is, consider intervals $\delta s_{n}$ of the normalized time $s$, then evolution of the system during the $n$-th time interval is supposed to be governed by a time-independent Hamiltonian given by $\hat{H}(\bar{s}_{n})$ evaluated at the medium point of the average point of the interval $\Scal_{n} = [\delta s_{n}, \delta s_{n+1}]$, namely $\bar{s}_{n}=(s_{n}+s_{n+1})/2$. In this way, we can write the evolution operator in this interval as
\begin{align}
	\hat{U}_{\mathrm{d}}(s_{n+1};s_{n}) \approx \exp\left[-i \left(f(\bar{s}_{n}) \hat{H}_{\mathrm{ini}} + g(\bar{s}_{n}) \hat{H}_{\mathrm{fin}}\right) \delta s_{n} \tau /\hbar\right] .
\end{align}

To understand the explicit $\tau$-dependence of $\hat{U}_{\mathrm{d}}(s_{n+1};s_{n})$, one should write the evolution operator in the time-domain $t$ as $\hat{U}_{\mathrm{d}}(t_{n+1};t_{n}) \approx \exp\left[-i \hat{H}(\bar{t}_{n}) \delta t_{n}  /\hbar\right]$, then change it to $s$-domain as follows $\hat{U}_{\mathrm{d}}(t_{n+1};t_{n}) \rightarrow \hat{U}_{\mathrm{d}}(s_{n+1};s_{n})$, by using $\delta t =  \tau \delta s$. In this way, we observe that the digitization procedure is $\tau$-dependent.

At this point we wonder the conditions on the system parameters such that the approximation in Eq.~\eqref{Eq:U_dig} is reached at first-order Trotter-Suzuki decomposition of the operator $\hat{U}_{\mathrm{d}}(s_{n+1};s_{n})$, with respect to the Hamiltonian in Eq.~\eqref{Eq:Hamiltonian}. In other words, for any adiabatic protocol, under which conditions can we have evolution operator of the form $\hat{U}_{\mathrm{d}}(s_{n+1};s_{n}) \approx \hat{\Ucal}(\hat{H}_{\mathrm{fin}}) \hat{\Ucal}(\hat{H}_{\mathrm{ini}})$,
where $\hat{\Ucal}(\hat{H}_{X})$ is unitary operator that implements an evolution driven by the Hamiltonian $\hat{H}_{X}$? Such a form for $\hat{U}_{\mathrm{d}}(s_{n+1};s_{n})$ is useful in context of adiabatic quantum optimization algorithms due to experimental feasibility~\cite{Bengtsson:20,Harrigan:21}, so we focus on this structure.

To properly address this question, first we use the Baker-Campbell-Hausdorff for the multiplication of two exponential functions of the Hamiltonians $\hat{H}_{\mathrm{ini}}$ and $\hat{H}_{\mathrm{fin}}$ as
\begin{align}
	e^{\beta \hat{H}_{\mathrm{fin} }}e^{\alpha \hat{H}_{\mathrm{ini}}}  =
	e^{ \alpha \hat{H}_{\mathrm{ini}} + \beta\hat{H}_{\mathrm{fin} }  + \frac{\alpha \beta}{2} \hat{A} + \frac{\alpha \beta^2}{12} \left[\hat{H}_{\mathrm{fin} } ,\hat{A}\right] - \frac{\alpha^2 \beta}{12} \left[\hat{H}_{\mathrm{ini} } ,\hat{A}\right] +\cdots } , \label{Eq:BCHExp}
\end{align}
with $\hat{A} = [\hat{H}_{\mathrm{fin} } ,\hat{H}_{\mathrm{ini}}]$, for two arbitrary $\alpha$ and $\beta$. Therefore, one realizes that by the suitable choice of $\alpha$ and $\beta$ as $\alpha = -i f(\bar{s}_{n}) \delta s_{n} \tau /\hbar$ and $\beta = -i g(\bar{s}_{n}) \delta s_{n} \tau /\hbar$, one gets
\begin{align}
	e^{- \frac{i\tau \delta s_{n}}{\hbar}g(\bar{s}_{n}) \hat{H}_{\mathrm{fin}} }e^{- \frac{i\tau \delta s_{n}}{\hbar}f(\bar{s}_{n}) \hat{H}_{\mathrm{ini}} } \approx
	e^{- \frac{i\tau \delta s_{n}}{\hbar}\hat{H}(\bar{s}_{n}) 
		+ \hat{\chi}(\bar{s}_{n}) + \Ocal\left((\delta s_{n})^4\right)} ,
\end{align}
where we defined
\begin{align}
	\hat{\chi}(\bar{s}_{n}) = g(\bar{s}_{n})f(\bar{s}_{n})\left[\frac{(i\tau \delta s_{n})^2}{2 \hbar^2} \hat{A} + \frac{(i\tau \delta s_{n})^3}{12 \hbar^3}\hat{B}(\bar{s}_{n})   \right],
\end{align}
with $\hat{B}(\bar{s}_{n}) =\big[g(\bar{s}_{n})\hat{H}_{\mathrm{fin} }- f(\bar{s}_{n})\hat{H}_{\mathrm{ini} } ,\hat{A}\big]$. Therefore, by choosing the total evolution time $\tau$ for the dynamics, from these equations we can state a first step to get a \textit{sufficient} condition to an efficient digital version of the adiabatic evolution. Initially, we need to satisfy the inequality (by neglecting terms of the order $(\delta s_{n})^4$)
\begin{align}
	\hbar	\norm{\hat{\chi}(\bar{s}_{n})} \ll \tau \delta s_{n}\norm{\hat{H}(\bar{s}_{n})} . \label{Eq:WeakCondition}
\end{align}

In principle, when we satisfy this inequality we are able to (approximately) mimic the dynamics driven by $\hat{H}(s)$ for a given total time $\tau$, which does not mean we have a perfect digitized approach for adiabatic approximation. In fact, the adiabatic regime is not obtained for any $\tau$, as suggested by the fundamental constraint on the adiabatic run time in Eq.~\eqref{Eq:AdTime}. Through this discussion, we are able to formulate fundamental conditions for the performance of digitized adiabaticity through the following theorem (see Appendix~\ref{ApTitle:TheoremProf} for further details).

\begin{theorem}\label{Theorem_Ad}
	Given $\hat{H}(s) = f(s) \hat{H}_{\mathrm{ini}} + g(s) \hat{H}_{\mathrm{fin}}$ an adiabatic Hamiltonian, a sufficient condition on the size step $\delta s_{n}$ of the digitized algorithm to mimic an adiabatic behavior, at first-order of the Suzuki-Trotter decomposition
	\begin{align}
		\hat{U}_{\mathrm{d}}(s_{n+1};s_{n}) \approx
		e^{- \frac{i\tau \delta s_{n}}{\hbar}g(\bar{s}_{n}) \hat{H}_{\mathrm{fin}} }e^{- \frac{i\tau \delta s_{n}}{\hbar}f(\bar{s}_{n}) \hat{H}_{\mathrm{ini}} } , \label{Eq:Approximation}
	\end{align}
	is given by
	\begin{align}
		\delta s_{n} \ll \delta s_{\mathrm{ad}} = \frac{1}{\tau_{\mathrm{ad}}}\min_{s} \left[\frac{1}{\norm*{\hat{B}(s)}}\left(\sqrt{\frac{\eta(s)}{f(s)g(s)} }
		-3 \norm{ \hat{A} }\right) \right] , \label{Eq:EqTheorem}
	\end{align}
	where $\eta(s) = 9 \norm*{ \hat{A} }^2 f(s)g(s)+12 \norm*{\hat{B}(s)} \norm*{\hat{H}(s)}$, 
		and the adiabatic time estimated as $\tau_{\mathrm{ad}} = \max_{0\leq s \leq 1}[\hbar\norm*{d_{s}\hat{H}(s)}/\Delta^2(s)]$.
\end{theorem}

The reference value $\delta s_{\mathrm{ad}}$ is named here as \textit{adiabatic digital resolution parameter}, since it provides an estimate about the maximal value for the time interval $\delta_{n}$ (resolution) of the discretized protocol. \revisionRefB{First of all, it is possible to see that the inequality in Eq.~\eqref{Eq:EqTheorem} requires the knowledge of physical and mathematical quantities that, in principle, may not be accessible in many-body systems from the experimental perspective. In such cases, the real value of the adiabatic digital resolution parameter used in experimental realizations may be estimated from the theoretical adiabatic Hamiltonian model.}

The above Theorem~\ref{Theorem_Ad} is the main result of this work. 
The above theorem establishes a condition to be satisfied to digitize the evolution in the limit under which the evolution is slowly-varying enough to reach the adiabatic dynamics regime. In other words, the size step $\delta s_{n}$ and the total evolution time $\tau$ are free parameters, but because the adiabatic evolution is only expected in the limit $\tau \gg \tau_{\mathrm{ad}}$, such a constraint has to be taken into account by the free parameter $\delta s_{n}$ if we want to perfectly mimic the adiabatic behavior. In fact, in case we want to digitize the evolution ruled by the time-dependent Hamiltonian $\hat{H}(s)$, at first order of the Suzuki-Trotter decomposition, the condition to be satisfied is given in Eq.~\eqref{Eq:WeakCondition}, which does not guarantee the adiabatic behavior. This discussion will be made clear when we consider some working examples in Sect.~\ref{Sec:Examples}.

As a consequence of this theorem, we can explain some results reported in Refs.~\cite{Barends:16,Mbeng:19}. First, we would like to highlight a relevant discussion reported in Ref.~\cite{Barends:16}. One of the empirical conclusions of~\cite{Barends:16} is the existence of \textit{intrinsic non-adiabatic} undesired transitions in the digital model, and these errors are not related to single and two-qubit gates imperfections. In this regard, the above theorem provides a mathematical demonstration of such ab empirical conclusion. In fact, the  Theorem~\ref{Theorem_Ad} can be used to establish the link between the analog (time-continuous) and digital non-adiabatic errors. To show that, let us consider that the total evolution time considered for the \textit{digital} implementation is $\tau_{\mathrm{dig}}$, if we take into account that the normalized time is defined as $s = t/\tau$, we can write for each time step $\delta s_{n} = \delta t_{n} / \tau_{\mathrm{dig}}$. Now, invoking the Theorem~\ref{Theorem_Ad} we have $\delta s_{n} \ll \delta s_{\mathrm{ad}}$, and it means that $\tau_{\mathrm{dig}} \gg \tau_{\mathrm{ad}}$, where $\tau_{\mathrm{ad}}$ is the adiabatic time provided by the validity conditions for adiabaticity~.

As for the result in Ref.~\cite{Mbeng:19}, through a benchmark of the adiabatic digitization of the Ising model the authors concluded that: \textit{given the total number of block $N_{\mathrm{bl}}$ of the digital circuit, there is an optimal choice $\tau_{\mathrm{opt}}$ of the total evolution time given by $\tau_{\mathrm{opt}} \sim N_{\mathrm{bl}}$}. To show that, first we recall that $\sum_{n=1}^{N_{\mathrm{bl}}} \delta s_{n} = 1$, so if we consider the case of a digitized algorithm with $\delta s_{n}$ given by the adiabatic parameter $\delta s_{\mathrm{ad}}$, we have $\sum_{n=1}^{N_{\mathrm{bl}}} \delta s_{\mathrm{ad}} = 1$, which leads to $\delta s_{\mathrm{ad}} = 1/N_{\mathrm{bl}}$. In addition, it is possible to rewrite the Theorem~\ref{Theorem_Ad} as $\delta s_{n} = \tau_{0} / \tau$~(see Appendix~\ref{ApTitle:TheoremProf}), for a given digitized adiabatic reference time $\tau_{0}$ (constant, independent on $N_{\mathrm{bl}}$). It leads to the conclusion that an optimal value of $\tau$ is expected to satisfy $\tau = \tau_{0} N_{\mathrm{bl}}$, or, similarly, $\tau \sim N_{\mathrm{bl}}$.

	To end, the condition established in this work provides a new and complementary perspective to the results presented in Ref.~\cite{Kovalsky:23}. In such a reference the authors claim that the error due to a first order Trotterization scaling with $\Ocal\big((\delta t/\tau)^2\big)$ may explain why the fidelity of DAQC enhances even for increased $\tau$. While our results corroborate to this conclusion, it introduces a new conditions on the size step $\delta s$. In fact, since the condition over $\delta s$ is always satisfied, even for a fixed $\delta s$, by increasing $\tau$ we can reach enhanced fidelity. However, when $\tau$ is too large, the value of $\delta s$ has to decrease in order to satisfy the Theorem~\ref{Theorem_Ad}, otherwise we induce errors due to the resolution of the digitization. In next section we exemplify all these predictions and conclusions through concrete applications.

\section{Digitizing adiabatic algorithms} \label{Sec:Examples}

In this section we present two example of applications of the aforementioned strategies and protocols. In particular, we will focus on two adiabatic algorithms: i) the two-qubit exact cover problem. as discussed in Ref.~\cite{Bengtsson:20},
and ii) the Factorization problem introduced by Peng \textit{et al.}~\cite{Peng:08}. The figure of merit considered here is the fidelity $\Fcal$ (or the infidelity, $\Ical$) between the output digital state $\ket{\psi_{\mathrm{out}}^{\mathrm{dig}}(N_{\mathrm{bl}})}$, after implementation of the circuit with $N_{\mathrm{bl}}$ blocks, and the analog adiabatic solution $\ket{\psi_{\mathrm{out}}^{\mathrm{ana}}}$. We assume $\ket{\psi_{\mathrm{out}}^{\mathrm{ana}}}$ as the solution of the Eq.~\eqref{Eq:Schrodinger} for the adiabatic Hamiltonian, with a total evolution time given by $\tau$, evaluated at the end of the evolution $s=1$ (similar to $t = \tau$). More precisely, consider the state $\ket{\psi_{\tau}(t)}$ as the solution of the Schrödinger equation driven by the adiabatic Hamiltonian for a total evolution time $\tau$. So, $\ket{\psi_{\mathrm{out}}^{\mathrm{ana}}}=\ket{\psi_{\tau}(t=\tau)}$. As measure of the digital error we define the \textit{infidelity} based on the Bures metric for pure states~\cite{Nielsen:06} as 
\begin{align}
	\Ical (N_{\mathrm{bl}}) = 1 - \Fcal (N_{\mathrm{bl}}) = 1 - | \interpro{\psi_{\mathrm{out}}^{\mathrm{dig}}(N_{\mathrm{bl}})}{\psi_{\mathrm{out}}^{\mathrm{ana}}} |^2 .
\end{align}

For a high fidelity implementation, $\Ical (N_{\mathrm{bl}}) \ll 1$.

\begin{figure*}[t!]
	\centering 
	\includegraphics[width=\linewidth]{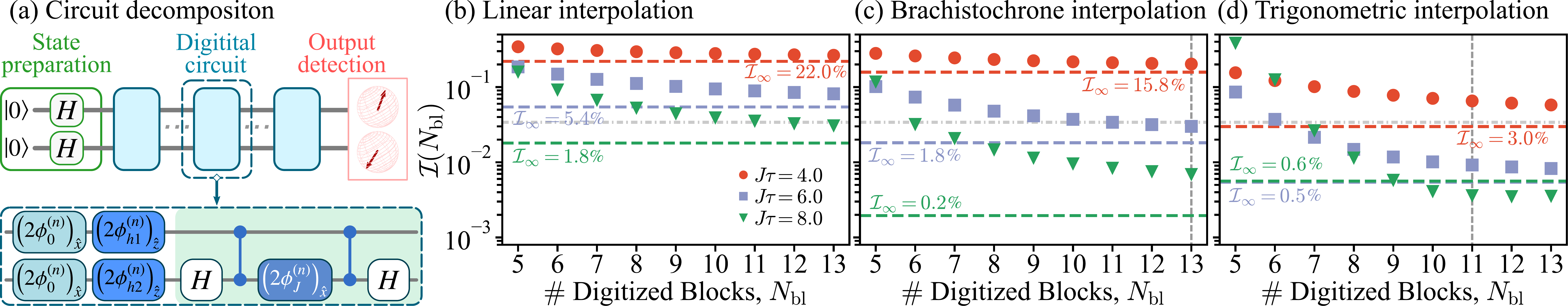}
	\caption{  Fidelity for digitized adiabatic evolution, for different values of $J \tau$, using the interpolation functions defined in Eqs.~\eqref{Eq:TrigoInter}, and~\eqref{Eq:LinearInter} and~\eqref{Eq:BrachInter}. In these graphs the total evolution time $\tau$ has been considered such that $J\tau = 4$, $J\tau = 6$ and $J\tau = 8$. To compare the performance of the models, the horizontal dashed lines denotes the expected result for the corresponding analog evolution. For sake of comparison, the horizontal dot-dashed line denotes the fidelity of $96.6\%$ achieved in Ref.~\cite{Bengtsson:20} using a double-layer QAOA. 
			Horizontal gray dashed lines are the minimum number of blocks $N_\mathrm{min} = 1/\delta s_\mathrm{ad}$ as predicted by the Theorem~\ref{Theorem_Ad}. For the linear case $N_\mathrm{min} = 17$, and for this reason it is now shown in the panel (b).
	}
	\label{Fig:GraphsAlgo}
\end{figure*}

\subsection{Two-qubits exact-cover problem} \label{Sec:ExactCover}

In order to exemplify the application of the optimal DQAC we consider the exact cover problem. As explained in Ref.~\cite{Bengtsson:20}, ``\emph{given a set $X$ and several subsets $\Scal_{i}$ containing parts of $X$, which combination of subsets include all elements of $X$ just once?}". A particular case of this task can be mapped into the two-qubit problem (final) Hamiltonian given by
\begin{align}
	\hat{H}_{\mathrm{fin}}/\hbar = h_1\hat{\sigma}_{1}^{z} + h_2\hat{\sigma}_{2}^{z} + J \hat{\sigma}_{1}^{z}\hat{\sigma}_{2}^{z} ,
\end{align}
with the initial Hamiltonian $\hat{H}_{\mathrm{ini}}/\hbar = h_{x} \left(\hat{\sigma}_{1}^{x}+\hat{\sigma}_{2}^{x}\right)$. The parameters $h_1$, $h_2$ and $J$ depends on the mathematical description of the problem. This dynamics is the simplest example of such a class of problems, but it is a timely example to address with our model. It has been implemented with a double-layer Quantum Approximate Optimization Algorithm (QAOA) with fidelity $96.6\%$ (over 200 iterations with random starting parameters)~\cite{Bengtsson:20}, and a single-layer QAOA the fidelity reported is around $50\%$. As stated by the authors in Ref.~\cite{Bengtsson:20}, the main complexity of this problem arises due to the classical optimization problem (due to the scaling of the space of parameters to be optimized). Here, we show that using a deterministic construction of the digitized algorithm, we can reach enhanced values of fidelity. To this end, the digitized version of this Hamiltonian is considered here for different total evolution times $\tau$ (according to the adiabatic theorem) and the interpolation functions $f$ and $g$, which allows for speeding the algorithm up (reduce $N_{\mathrm{bl}}$).

To obtain the circuit parameters we need to set the total evolution time $\tau$ (we choose dimensionless values for $J\tau$) and the total number of Blocks $N_{\mathrm{bl}}$. So, using the Eq.~\eqref{Eq:Approximation} we find the unitary for the $n$-th block of the exact cover problem as
\begin{align}
	\hat{U}_{\mathrm{d}}^{\mathrm{ExCo}}(s_{n+1};s_{n}) =
	e^{- i\phi_{J}^{(n)} \hat{\sigma}_{1}^{z}\hat{\sigma}_{2}^{z} }e^{- i(\phi_{h1}^{(n)}\hat{\sigma}_{1}^{z} + \phi_{h2}^{(n)}\hat{\sigma}_{2}^{z}) } e^{- i\phi_{0}^{(n)} ( \hat{\sigma}_{1}^{x}+\hat{\sigma}_{2}^{x} ) },
\end{align}
where we already used that $[h_1\hat{\sigma}_{1}^{z} + h_2\hat{\sigma}_{2}^{z}, J \hat{\sigma}_{1}^{z}\hat{\sigma}_{2}^{z}]=0$, and
\begin{align}
	\phi_{0}^{(n)} = \tau  h_{x} f(\bar{s}_{n})\delta s_{n}, ~ \phi_{hk}^{(n)} = \tau h_k g(\bar{s}_{n})\delta s_{n}, ~ \phi_{J}^{(n)} = \tau  J \delta s_{n}g(\bar{s}_{n}),
\end{align}
as the rotation angles of the gates that implements the initial local $\hat{\sigma}_{x}$ terms, the final local $\hat{\sigma}_{z}$ terms, and the ZZ-interaction, respectively. The complete circuit that implements the set of unitary above is as shown in Fig.~\SubFig{Fig:GraphsAlgo}{a}. The circuit is decomposed using the single qubit rotations of a angle $\theta$ around the directions $\xi=\{x,y,z\}$ denoted as $(\theta)_{\xi} = \hat{R}_{\xi}(\theta) = e^{- i\theta \hat{\sigma}_{\xi}/2}$~\cite{Nielsen:Book}, with $\hat{\sigma}_{\xi} = \{\hat{\sigma}_{x},\hat{\sigma}_{y},\hat{\sigma}_{z}\}$ the Pauli matrices for a single qubit. After the state preparation step, we apply $N_{\mathrm{bl}}$ blocks with step-dependent parameters, where each block is constituted of single and two-qubit CZ gates (first and second lines in the circuit denote qubit 1 and 2, respectively). The two first set of gates in each digital block (highlighted in the dashed line box) implement the local fields $\hat{\sigma}_{x}$ and $\hat{\sigma}_{y}$ respectively, while the $ZZ$-interaction is simulated by CZ gates with help of Hadamard gates and a single-qubit rotation in $\hat{x}$. Each parameter of the circuit is obtained after setting the parameters of the Hamiltonian, where our first choice is $h_x = J$ (no generality loss). The exact cover is mapped into this Hamiltonian of $h_1$, $h_2$ and $J$, such that here we choose the example in which one has two sets $S_1 =\{z_{1}, z_{2}\}$ and $S_2 =\{z_{1}\}$. For this particular case, the adiabatic solution reads $\ket{\psi_{\mathrm{ad}}} = \ket{1}\ket{0}$, since the repeated element is $z_{1}$, and the problem Hamiltonian has $h_1=J$, $h_2=0$. 

\begin{figure*}[t!]
	\centering
	\includegraphics[width=1.0\linewidth]{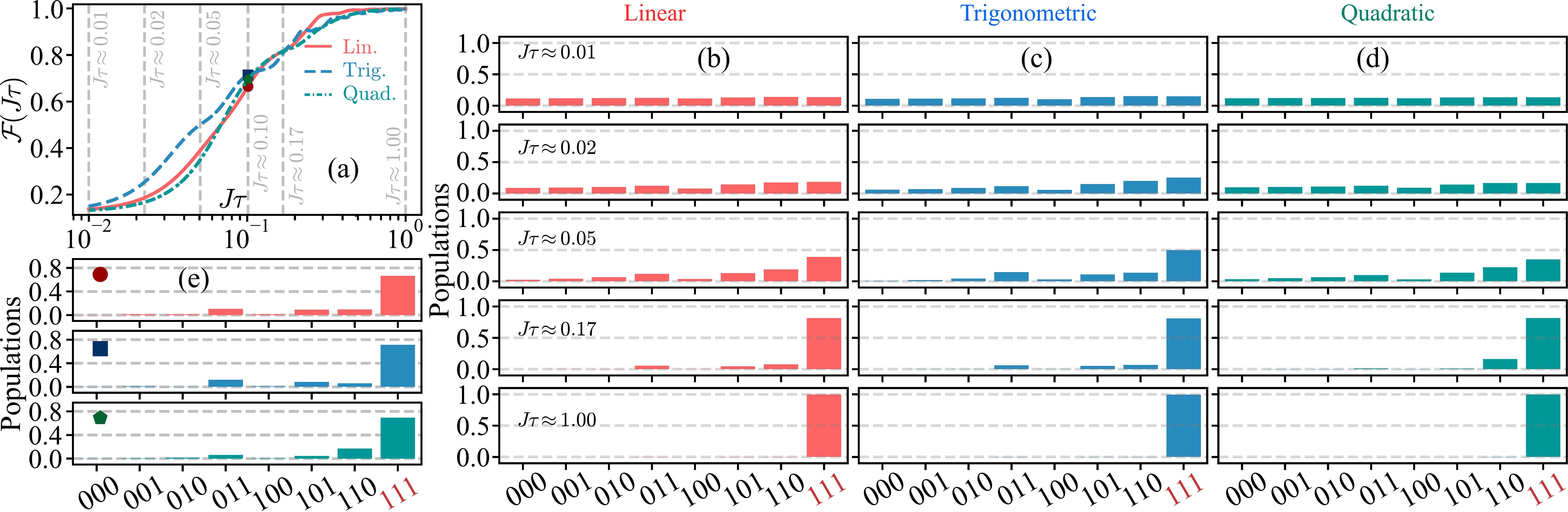}
	\caption{(a) Fidelity of the analog adiabatic solution with respect to the for different choices of the interpolation. (b--d) populations for the analog (time-continuous) adiabatic Shor's algorithm for different values of the total run time $\tau$.  In our simulation we considered $h_{x} = 30$.}
	\label{Fig:Factorization}
\end{figure*}

To end, we now set the interpolation functions $f(s)$ and $g(s)$ by assuming three different cases:
\begin{align}
	f_{\mathrm{T}}(s) &= \cos(s \pi / 2) , ~~ g_{\mathrm{T}}(s) = \sin(s \pi / 2) \label{Eq:TrigoInter} \\
	f_{\mathrm{L}}(s) &= 1 - s , ~~ g_{\mathrm{L}}(s) = s \label{Eq:LinearInter} \\
	f_{\mathrm{B}}(s) &= \frac{1}{2} + \frac{1}{2}\tan\left(\frac{(1-2s)\pi}{4} \right) , ~~ g_{\mathrm{B}}(s) = 1 - f_{\mathrm{B}}(s) \label{Eq:BrachInter},
\end{align}
called here trigonometric, linear and brachistochrone interpolation functions, respectively. These three functions have been considered here for comparison purposes, since each of those provides a different adiabatic run time $\tau_{\mathrm{ad}}$ due to the difference in the Hamiltonian minimum energy gap. In fact, by computing the adiabatic time using Eq.~\eqref{Eq:AdTime} we find the values $|J\tau_{\mathrm{ad}}^{\mathrm{L}}|= 4.0$, $|J\tau_{\mathrm{ad}}^{\mathrm{T}}| \approx 2.2$ and $|J\tau_{\mathrm{ad}}^{\mathrm{B}}|= \pi$, for the trigonometric, linear and brachistochrone functions, respectively. These values give us the ideal scenario to discuss the result claimed by the Theorem~\ref{Theorem_Ad}, where the optimal solution will be only found when we satisfy its condition. In order to observe how the performance of the circuit behaves as function of the number of blocks and our choice on the total run time $\tau$, we consider the cases where $|J\tau| = \{4, 6 , 8\}$. For simplicity, here we consider the case in which the Trotterization is done for identical time steps $\delta s_{n} = \delta s_{0} = 1 / N_{\mathrm{bl}}$.

In Figs.~\SubFig{Fig:GraphsAlgo}{b}--\SubFig{Fig:GraphsAlgo}{d} we present the result for the infidelity (error) $\Ical (N_{\mathrm{bl}})$ of the protocol in finding the correct adiabatic output for all interpolation functions defined in Eqs.~\eqref{Eq:TrigoInter}--\eqref{Eq:BrachInter}. Each set of points denote the behavior of $\Ical (N_{\mathrm{bl}})$ for a given total run time $|J\tau|$ as function of the circuit depth $N_{\mathrm{bl}}$. For sake of comparison, we also show the expected infidelity of the output state obtained by the analog adiabatic evolution as horizontal dashed lines, for each choice of the total evolution time and interpolation function. The simulation corroborates with the relations of the adiabatic total evolution time $|J\tau_{\mathrm{ad}}^{\mathrm{T}}|<|J\tau_{\mathrm{ad}}^{\mathrm{B}}|<|J\tau_{\mathrm{ad}}^{\mathrm{L}}|$, such that the trigonometric function develops the best performance for all choices of the run time $|J\tau|$. This example is a timely showcase to reinforce our claim that high performance digitized quantum annealers have to take into account the adiabatic theorem and its traditional gap condition~\cite{Sarandy:04,Amin:09}. The behavior of the digitized infidelity approaches the horizontal lines as $N_{\mathrm{bl}}$ increases, and one expects to reach such asymptotic infidelities for a infinite-sized digital protocol ($N_{\mathrm{bl}}\rightarrow\infty$). For the dynamics under consideration, it is intuitive to observe that digitized algorithms are mostly worse than their analog counterparts. However, one can find cases in which the digital annealing achieves fidelities better than its analog corresponding (Fig.~\SubFig{Fig:GraphsAlgo}{d} for $|J\tau| = 8.0$).

	As aforementioned, we observe a non-trivial behavior for the infidelity which depends on the interpolation functions, total evolution time and resolution of the digital algorithm (size step $\delta s$). In particular, for $\delta s$ smaller enough as stated by the Theorem~\ref{Theorem_Ad}, our results are in agreement with previous predictions~\cite{Barends:16,Mbeng:19,Kovalsky:23}. As we shall see now, such a counterintuitive performance of the digital circuit can also be observed for more complex algorithms.

\subsection{Adiabatic Factorization}

We now discuss how to efficiently implement the quantum factorization~\cite{Shor:94}, through its adiabatic version~\cite{Peng:08}. More precisely, we shall focus on the factorization of the number $21$. Such an adiabatic algorithm aims at minimizing the quadratic cost function given as $f(q,p) = (N - qp)^2$, where its minimum in the parameter space $(q,p)$ is zero and it is achieved when $N = q\times p$. In other words, by minimizing $f(q,p)$ we find the integer number $q,p$ which factorize $N$. The information of $q,p$ is encoded in the computational basis $\ket*{x_{1}x_{2}\cdots x_{N_\mathrm{qu}}}$, with $x_{1}\in \{0,1\}$. In particular, for the case of the number $21$, such an algorithm requires $N_\mathrm{qu}= 3$ qubits in which the solution reads $p = 2 x_{1} + 1$ and $q = 4 x_{2} + 2x_{3} + 1$. To start with, it has been shown that the adiabatic problem Hamiltonian for the factorization is
\begin{align}
	\hat{H}^{\mathrm{fact}}_{\mathrm{fin}} =J\hbar \sum_{k=1}^{3} h_{k}\sigma_{z}^{k} + \sum_{(k,m)} j_{k,m}\sigma_{z}^{k} \sigma_{z}^{m} + j_{3} \sigma_{z}^{1}\sigma_{z}^{2} \sigma_{z}^{3} ,
\end{align}
in which the sum over $(k,m) = \{(1,2),(2,3),(1,3)\}$. Here, we assume the dimensionless parameters $h_{1} = 84$, $h_{2} = 2 h_{3} = 88$, $j_{1,2} = -2j_{1,3} = -j_{2,3} = -20$, and $j_{3}=- 16$. It is worth mentioning that an additional term in $\hat{H}_{\mathrm{P}}$ is used in Ref.~\cite{Peng:08}, but because it only implies an energy shift of the entire energy spectrum, it does not affect the adiabatic dynamics and its validity conditions. The protocol states that the three-qubit system should be driven by the time-dependent Hamiltonian in Eq.~\eqref{Eq:Hamiltonian}, with the initial Hamiltonian $H^{\mathrm{fact}}_{\mathrm{ini}} = h_{x} \sum_{n=1}^{3} \hat{\sigma}_{x}^{n}$, where we choose (without loss of generality) $h_{x}>0$ which leads to the initial ground state $\ket{\psi(0)} = \ket{---}$, where we define $\ket{\pm}=(\ket{0}\pm\ket{1})/\sqrt{2}$. 

\begin{figure}[t!]
	\centering
	\includegraphics[width=\columnwidth]{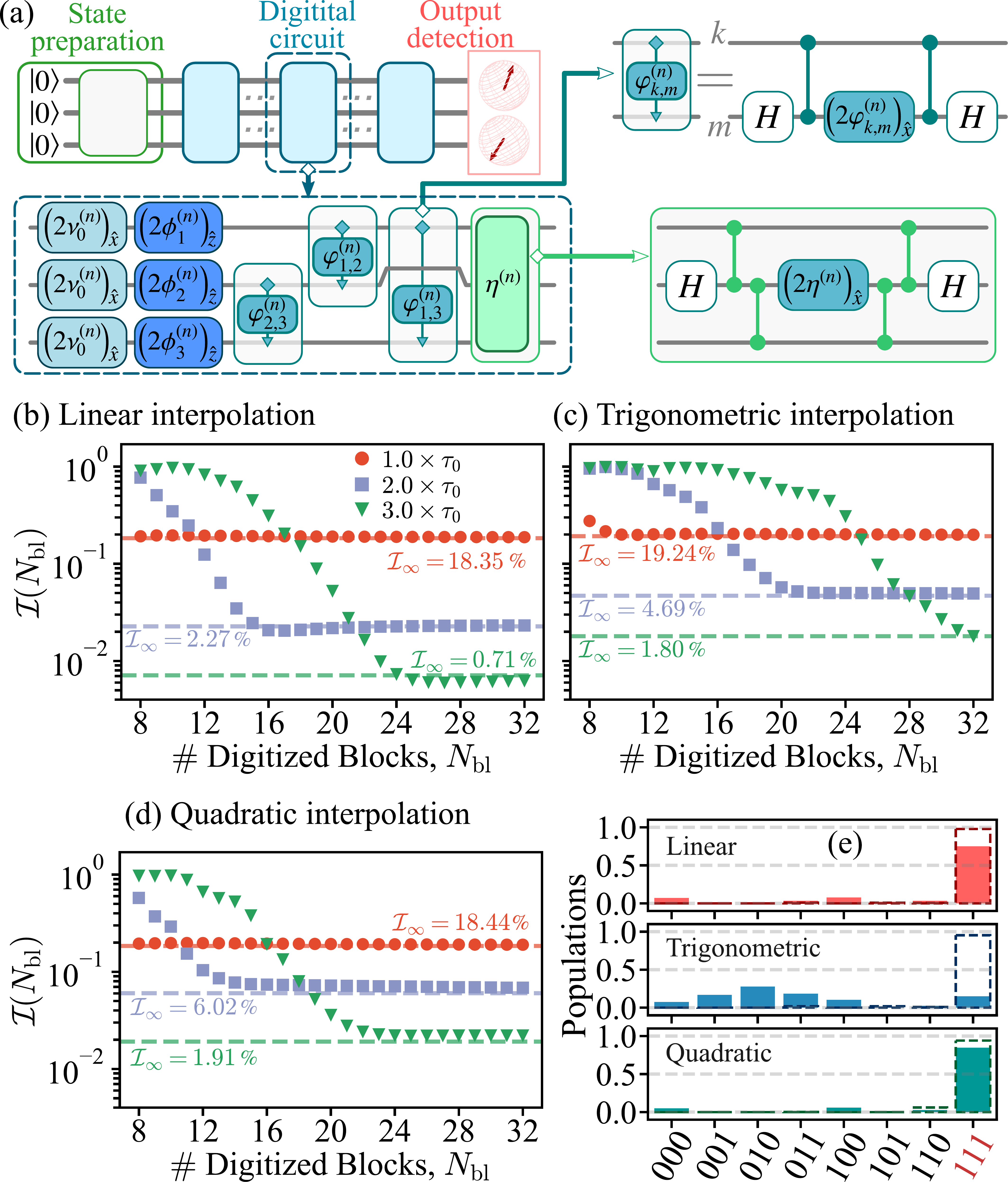}
	\caption{(a) Circuit for the digitized adiabatic \revisionRefB{factorization} algorithm. (b--d) Infidelity as function of the number $N_{\mathrm{bl}}$ for different choices of $\tau$ and interpolation functions. Horizontal dashed lines are the infidelity of the corresponding adiabatic state with respect to the solution of the problem $\ket{111}$. (e) Output state population (in the computational basis) for the digitized algorithm when $\tau = 2\tau_{0}$ and $N_{\mathrm{bl}} = 11$. The parameters of the Hamiltonian are the same as in Fig.~\ref{Fig:Factorization}. 
			For all cases considered here the minimum number of blocks are much bigger than $N_\mathrm{min} = 32$, which reinforces that our condition is \textit{sufficient}, but it is not \textit{necessary}. \revisionRefB{The circuit in the panel (a) is not the Shor's algorithm circuit, which means we are not adiabatically implementing the Shor's algorithm but rather an alternative approach to factorization.}
	}
	\label{Fig:FactorizationDigitized}
\end{figure}

By considering different interpolation functions and total evolution times we can now discuss the preliminary analysis from the results of the Fig.~\ref{Fig:Factorization}. In this algorithm we also considered the trigonometric and linear interpolations already defined in Eqs.~\eqref{Eq:TrigoInter} and~\eqref{Eq:LinearInter}, in addition to the quadratic interpolation used in Ref.~\cite{Peng:08} and defined as
\begin{align}
	f_{\mathrm{Q}}(s) = 1 - s^2 , ~~ g_{\mathrm{Q}}(s) = s^2 \label{Eq:QuadraticInter} .
\end{align}

First, from the behavior of the corresponding fidelities for each interpolation function shown Fig.~\SubFig{Fig:Factorization}{a}, one identifies that this algorithm can be adiabatically implemented with high fidelity in the regime of evolution such that $|J \tau| \lessapprox 1.0$ for any interpolation considered here. In particular, we pay attention to the linear interpolation, which can reach the highest values of fidelity faster than the trigonometric and quadratic ones. However, it is possible to observe that for an evolution speed of $|J\tau| \approx 0.1$, all  interpolation functions already provide a fidelity around  $65\%-70\%$ of recognizing the correct solution of the problem (namely, the state $\ket{111}$). This fidelity can be even better if we drive the system at a total evolution time $\tau_{0} \approx 0.17 / |J|$, as done in the experiment of Ref.~\cite{Peng:08}. In Figs.~\SubFig{Fig:Factorization}{b}--\SubFig{Fig:Factorization}{d} we can observe the emergence of the correct solution with respect to the computational basis encoding all possible outputs. To this end, for each choice of the interpolation function, we adiabatically drive the system during a given total evolution time $|J\tau|$, and then we measure the populations of the density matrix in the computational basis $\ket{x_1 x_2 x_3}$, where $x_k = \{0,1\}$. In particular, we observe that the choice of the total evolution time as $\tau = \tau_{0}$ allows us to achieve the desired solution. In addition, as shown in Fig.~\SubFig{Fig:Factorization}{e}, we also could consider $|J\tau| \approx 0.1$, since the population in the state $\ket{x_1 x_2 x_3}$ allows us to identify the solution. However, we will use $\tau_{0}$ as the reference time in our analysis of the digitized implementation due to the fidelity associated to this choice. 

By following the same procedure as before, the digitization of the adiabatic Shor algorithm can be done as follows. First, we observe that the Suzuki-Trotter decomposition (in first order) of the adiabatic evolution operator provides
\begin{align}
	\hat{U}_{\mathrm{d}}^{\mathrm{fact}}(s_{n+1};s_{n}) = \hat{U}_{\mathrm{d}}^{\mathrm{int}}(s_{n+1};s_{n}) e^{- i\sum\nolimits_{k=1}^{3}\phi_{k}^{(n)}\hat{\sigma}_{k}^{z}  } e^{- i\nu_{0}^{(n)} ( \hat{\sigma}_{1}^{x}+\hat{\sigma}_{2}^{x} ) },
\end{align}
where the interaction part given by
\begin{align}
	\hat{U}_{\mathrm{d}}^{\mathrm{int}}(s_{n+1};s_{n}) = e^{- i\eta^{(n)} \hat{\sigma}_{1}^{z}\hat{\sigma}_{2}^{z}\hat{\sigma}_{3}^{z} } e^{- i\sum\nolimits_{(k,m)}\varphi_{k,m}^{(n)}\hat{\sigma}_{k}^{z}\hat{\sigma}_{m}^{z}  } , 
\end{align}
where we defined the dimensionless parameters
\begin{align}
	\nu_{x}^{(n)} &= \tau h_{x} f(\bar{s}_{n}) \delta s_{n} ,~~ \phi_{k}^{(n)} = \tau J  h_{k} g(\bar{s}_{n})\delta s_{n} , \\
	\varphi_{k,m}^{(n)} &= \tau J   j_{k,m}g(\bar{s}_{n})\delta s_{n} ,~~ \eta^{(n)} = \tau J j_{3}g(\bar{s}_{n}) \delta s_{n} ,
\end{align}
where again we shall consider the simplest case in which we use identical time steps $\delta s_{n} = \delta s_{0} = 1/N_{\mathrm{bl}}$ for the Trotterization. The circuit that implements the evolution given by $\hat{U}_{\mathrm{d}}^{\mathrm{fact}}(s_{n+1};s_{n})$, for the three qubit system, is shown in Fig.~\ref{Fig:FactorizationDigitized}. As done previous, we considered a quantum processor in which CZ gates can be implemented in a native (single-shot) way.

The Figs.~\SubFig{Fig:FactorizationDigitized}{b}--\SubFig{Fig:FactorizationDigitized}{d} show the infidelity, as function of $N_{\mathrm{bl}}$, of achieving the correct answer of the algorithm for each interpolation function used. In all the cases considered here, the results for the case $\tau = 1.0\tau_{0}$ strongly corroborate to the previous claim that the efficiency of the digitized algorithm is also bounded by the performance of the corresponding adiabatic evolution. From Figs.~\SubFig{Fig:FactorizationDigitized}{b}--\SubFig{Fig:FactorizationDigitized}{d}, we take a look at the particular choice of the total evolution time $\tau = 2\tau_{0}$ and number of blocks $N_{\mathrm{bl}} = 11$. Although the linear achieve better asymptotic fidelities, under this choose of parameters we can speed up the digitized implementation of the algorithm with enhanced fidelities using the quadratic interpolation function. In fact, from the population distribution in the computational basis shown in Fig.~\SubFig{Fig:FactorizationDigitized}{e} we can observe the performance in identifying the correct answer for the three interpolation functions, where we also include the comparison with their corresponding adiabatic case (dashed bar plots). We stress here the fact that the linear and quadratic interpolation provide a clear identification of the algorithm solution, both in the digitized and analog implementation of the algorithm. However, the digitized adiabatic version for trigonometric interpolation does not provide any indicative of the correct answer for the problem, even when its analog version develops high performance on the prediction of the correct target state. It reinforces the advantage of considering a previous analysis on the choice of the interpolation function before the digitized annealing protocol.

\revisionRefB{This analysis is pivotal to reinforce the Theorem~\ref{Theorem_Ad} as a condition to get, with good fidelity, the same solution as provided by the adiabatic algorithm. It does not mean that the solution of the problem needs a high fidelity protocol to be found, as it depends on how efficient the analog adiabatic algorithm is. For example, in Fig.~\SubFig{Fig:FactorizationDigitized}{e} we can see that the digital model, with linear interpolation, can be used to efficiently identify the solution of the problem, even when the corresponding fidelity with respect to the adiabatic solution is around $80\%$ (see Fig.~\SubFig{Fig:FactorizationDigitized}{b} for $\tau = 2\tau_0$ and $N_\mathrm{nb}=11$). It is particularly possible because the adiabatic algorithm is able to find the correct answer without a high fidelity evolution.}

For sake of completeness, we would like to highlight an additional discussion with respect to the experiment by Peng \textit{et al}~\cite{Peng:08}. It is interesting to mention that the Trotterization considered by Peng \textit{et al} is of the form
\begin{align}
	\hat{U}_{\mathrm{Peng}}^{\mathrm{fact}}(s_{n+1};s_{n}) = e^{- \frac{i\nu_{x}^{(n)}}{2} \sum_{k=1}^{2}\hat{\sigma}_{k}^{x} } \hat{U}_{\mathrm{d}}^{\mathrm{fact}}(s_{n+1};s_{n}) e^{ \frac{i\nu_{x}^{(n)}}{2} \sum_{k=1}^{2}\hat{\sigma}_{k}^{x} } .
\end{align}

\begin{figure}[t!]
	\centering
	\includegraphics[width=\linewidth]{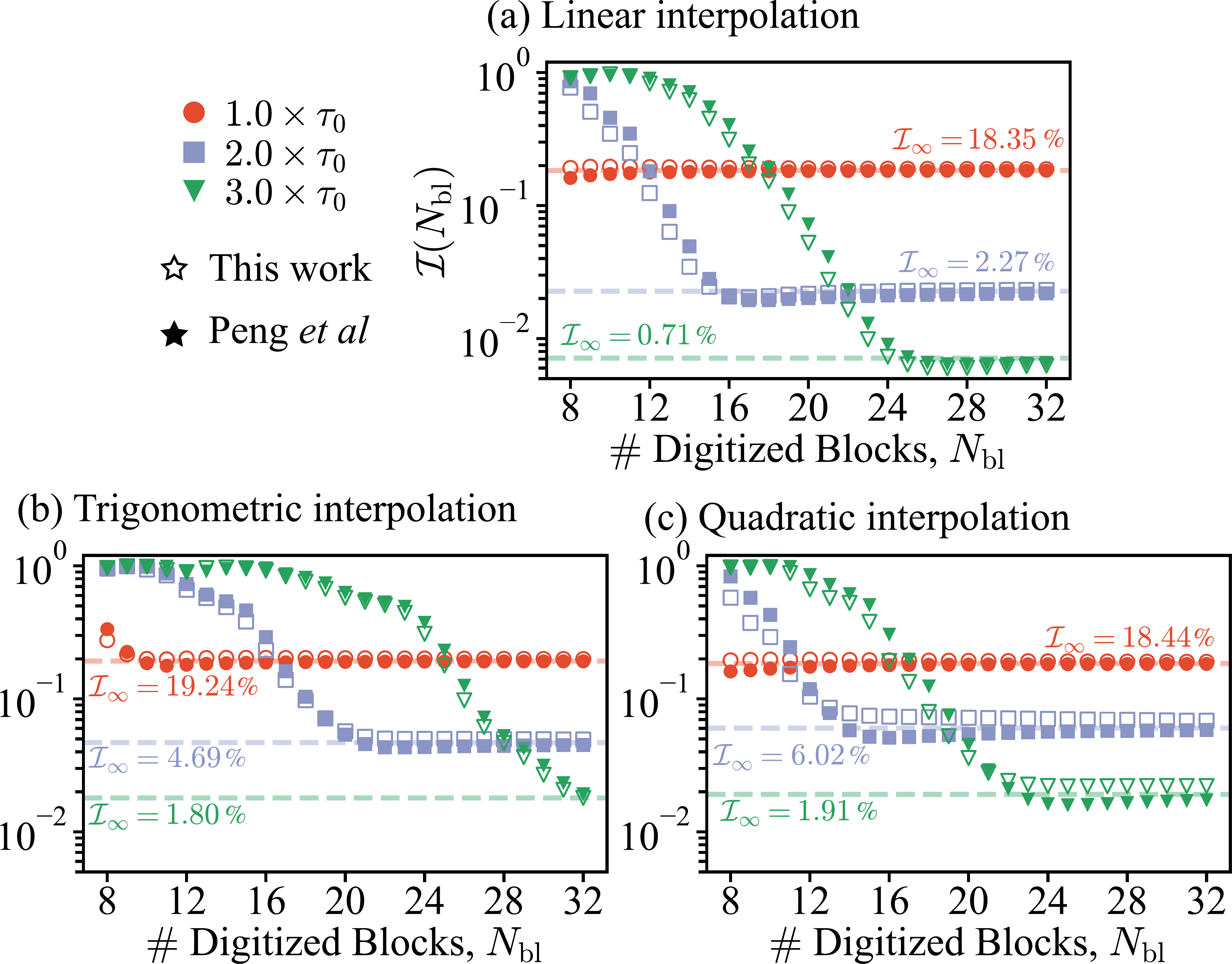}
	\caption{Comparison between the circuit decomposition considered in this work (open symbols) and the choice done in Ref.~\cite{Peng:08} (fulled symbols) for the three different interpolations, as considered in Fig.~\ref{Fig:FactorizationDigitized}, and for different choices of the total evolution time as a multiple of $\tau_0$. The Hamiltonian parameters are the same as in Fig.~\ref{Fig:Factorization}.}
	\label{Fig:Peng_Digital_x_Analog}
\end{figure}

Under this choice, the initial local operations that simulate the initial Hamiltonian is decomposed as a ``sandwiched ordering" of the operations. This choice does lead to reinterpretation of  adiabatic protocol, therefore we remark that such a way for the Trotter-Suziki decomposition of the evolution operator is smart option (at least for the algorithm under study here). In fact, in Fig.~\ref{Fig:Peng_Digital_x_Analog} we compare the ``normal ordering" proposed in this work and the ``sandwiched ordering" by Peng \textit{et al}~\cite{Peng:08}. It is possible to observe regimes in which the ``sandwiched" case leads to a slightly enhanced fidelity of the protocol. However, for the quadratic interpolation (used in their experiment) the performance of the ``sandwiched ordering" becomes notable, reaching values better than its analog counterpart. We understand this result as an indicator that the condition introduced in Theorem~\ref{Theorem_Ad} may be sensitive to the kind of Trotter decomposition for the digitized processing.

\section{Complexity of Adiabatic Digitization}

The discussion done so far focus on the complexity and scaling of the digitization process taking into account the conditions for adiabaticity. However, the effective complexity of the DAQC goes beyond this aspect, since the realization of the evolution operators for the initial $\hat{H}_{\mathrm{ini}}$ and final Hamiltonian $\hat{H}_{\mathrm{fin}}$, in Eq.~\eqref{Eq:Approximation}, should be taken into account. In fact, as it is possible to observe from the examples considered here, the exact-cover problem and factorization are implemented with the problem Hamiltonian given by the Ising ZZ Hamiltonian, but the factorization requires a three-body term that needs to decomposed in two-qubit gates. Consequently, the circuit to the digitized factorization demands more operations than the exact-cover circuit due to the three-body term $\hat{\sigma}_{1}\hat{\sigma}_{2}\hat{\sigma}_{3}$`. Therefore, by considering the decomposition of $\hat{H}_{\mathrm{ini}}$ and $\hat{H}_{\mathrm{fin}}$, into fundamental single- and two-qubit gates, we can estimate the algorithm complexity for adiabatic digitization. In particular, because $\hat{H}_{\mathrm{ini}}$ contains only single-qubit local fields in mostly of the adiabatic algorithms, let us focus on the implementation of $\hat{H}_{\mathrm{fin}}$. 

For an arbitrary $\hat{H}_{\mathrm{fin}}$, we can \textit{estimate} the circuit depth to simulate each block of the digital circuit using the complexity to decompose an arbitrary unitary gate for $N$ qubits, as $e^{- \frac{i\tau \delta s_{n}}{\hbar}g(\bar{s}_{n}) \hat{H}_{\mathrm{fin}} }$ can be interpreted as a $N$-qubits unitary. In this case, the complexity will scales significantly depending on the kind of gate available in the digital simulator and the number of $k$-local terms of $\hat{H}_{\mathrm{fin}}$. In particular, for Hamiltonian $\hat{H}_{\mathrm{fin}}$ with $k$-local terms, the simulation of an evolution of the form $e^{- i \hat{H} \xi}$ scales with $N^{k} \xi ||\hat{H}||_{1}$, with $||\hat{H}||_{1} = \sum_{\ell} ||\hat{h}_\ell||$, with $\hat{h}_\ell$ the $\ell$-th $k$-body term of $\hat{H}$~\cite{Low:19,Childs:21}. Regardless the digital algorithm used to simulate Hamiltonians, like QAOA or the digitized annealing, this complexity analysis is universal because it is related to decomposition of the $e^{- i \hat{H} \xi}$. 

For the cases considered here, as ``trivial" cases, we exploited the fact that whenever the problem Hamiltonian $\hat{H}_{\mathrm{fin}}$ admits only interaction between qubits that are physically connected in the quantum hardware, the additional complexity scales with $\Ocal(N_{\mathrm{int}})$, with $N_{\mathrm{int}}$ is the number of interaction terms in $\hat{H}_{\mathrm{fin}}$, as we can use the physical connectivity of the device to implement the necessary operations. For example, consider the simulation of a linear spin chain where $\hat{H}_{\mathrm{fin}}$ has nearest-neighbor $ZZ$ interactions. In this case, one expects the complexity to digitize this Hamiltonian in a linear chain of qubits, through single-qubit rotations and controlled $Z$ gates, scales with $\Ocal(N)$, with $N$ the number of qubits. 

\section{Conclusions and prospects}

In this work we introduced a validity condition for high-fidelity digitized quantum annealing, in analogy to validity conditions for adiabaticity exploited in literature~\cite{Sarandy:04,Amin:09,Jansen:07,Cao:13,Lidar:09,Messiah:Book}.  To illustrate the relevance of the results obtained in this study, we showed how the Theorem~\ref{Theorem_Ad} allows us to explain experimental and numerical results related to the efficiency of DAQC in different contexts~\cite{Barends:16,Mbeng:19}. Also, as an immediate consequence of the approach used here to demonstrate the Theorem~\ref{Theorem_Ad}, we conclude that making the number of blocks of the circuit large does not imply we will achieve high fidelity DAQC. In fact, the performance of a digital annealer depends on both the total number of blocks $N_{\mathrm{bl}}$ and the fundamental constraint of the adiabatic theorem related to the adiabatic time $\tau_{\mathrm{ad}}$. This result was exemplified by digitizing the adiabatic version of the two-qubits exact-cover problem and the adiabatic factorization. 

It is import to emphasize that DAQC is a potential candidate alternatively to QAOA due two properties of this model. First, through the examples exploited here one concludes that we do not need a classical optimizer after each experimental implementation of the circuit layers. This constitutes the first relevant difference with respect to QAOA. This leads us to the second property, which is related to the robustness of the solution of the circuit of our model with respect to any exponential complexity observed for QAOA algorithms when the number of layers is bigger than 2~\cite{Farhi:14,Bengtsson:20}.
Each model has its own set of advantages, but it is not possible to state a global advantage of one method with respect to the other one. For this reason, at this stage of development we can only state that DAQC can be used in scenarios to which QAOA may not be efficient~\cite{Blekos:24}. For example, when the number of independent parameters is too large, such that the classical optimization stage of QAOA is not able to quickly converge to a solution.

As future prospects on the subject of this work, it is worth saying that the Theorem~\ref{Theorem_Ad} is supported by a \textit{sufficient}, but not \textit{necessary}, validity condition to adiabaticity~\cite{Sarandy:04,Amin:09,Jansen:07,Cao:13,Lidar:09}. For this reason, the generalizations and further investigations on more precise conditions for digitized annealing supported by sufficient \textit{and} necessary adiabaticity conditions~\cite{Tong:05,Comparat:09,Suter:08,Hu:19-b} are desired. Also, it is worth highlighting that all examples used in this study we considered the case in which the incremental intervals $d s_{n}$ are identical, i.e., $d s_{n}=d s_{0}=1/N_{\mathrm{bl}}$. It means that further optimization can be made by fixing the number of total blocks $N_{\mathrm{bl}}$ and finding the distribution of intervals $\{ d s_{n} \}$, satisfying $\sum_{n=1}^{N_{\mathrm{bl}}} d s_{n} = 1$, to enhance our results discussed here. It can be done using Machine Learning techniques, for example. Also, it is timely to stress that the discussion on the complexity done in this work has to be understood as an \textit{estimate} of the number of gates to implement an arbitrary digital annealing process, as its accurate complexity remains as an open question to be addressed.

\revisionRefB{The strategies and results presented in this work have potential of applications to different approaches of quantum annealing. For example, by relaxing the conditions and requirements to adiabaticity considered in this work, in principle, one can develop a new hybrid digitized version of Diabatic Quantum Annealing and Diabatic Quantum Computation~\cite{Somma:12,Muthukrishnan:16,Crosson:21,Cote:23,Gerblich:24,Banks:24}, opening a new avenue to applications beyond digitized adiabatic tasks.}

\begin{acknowledgments}
The author is supported by the Comunidad de Madrid through the research funding program Talento 2024 ``César` Nombela", under the grant No 2024-T1/COM-31530.
The author acknowledges the partial support by the European Union's Horizon 2020 FET-Open project SuperQuLAN (899354), and by Proyecto Sinérgico CAM 2020 Y2020/TCS-6545 (NanoQuCo-CM) from the Comunidad de Madrid.
\end{acknowledgments}

\appendix


\onecolumngrid

\section{Proof of the Theorem~\ref{Theorem_Ad}}
\label{ApTitle:TheoremProf}

The starting point of our proof is the Baker-Campbell-Hausdorff for the multiplication of two exponential functions of the Hamiltonians $\hat{H}_{\mathrm{ini}}$ and $\hat{H}_{\mathrm{fin}}$ as
\begin{align}
	e^{\beta \hat{H}_{\mathrm{fin} }}e^{\alpha \hat{H}_{\mathrm{ini}}}  =
	e^{ \alpha \hat{H}_{\mathrm{ini}} + \beta\hat{H}_{\mathrm{fin} }  + \frac{\alpha \beta}{2} [\hat{H}_{\mathrm{fin} } ,\hat{H}_{\mathrm{ini}}] + \frac{\alpha \beta^2}{12} \left[\hat{H}_{\mathrm{fin} } ,[\hat{H}_{\mathrm{fin} } ,\hat{H}_{\mathrm{ini}}]\right] - \frac{\alpha^2 \beta}{12} \left[\hat{H}_{\mathrm{ini} } ,[\hat{H}_{\mathrm{fin} } ,\hat{H}_{\mathrm{ini}}]\right] +\cdots } .
\end{align}

As mentioned in the main text, we can choose $\alpha = -i f(\bar{s}_{n}) \delta s_{n} \tau /\hbar$ and $\beta = -i g(\bar{s}_{n}) \delta s_{n} \tau /\hbar$
\begin{align}
	e^{- \frac{i\tau \delta s_{n}}{\hbar}g(\bar{s}_{n}) \hat{H}_{\mathrm{fin}} }e^{- \frac{i\tau \delta s_{n}}{\hbar}f(\bar{s}_{n}) \hat{H}_{\mathrm{ini}} } \approx
	e^{- \frac{i\tau \delta s_{n}}{\hbar}\hat{H}(\bar{s}_{n}) 
		+ \hat{\chi}(\bar{s}_{n}) + \Ocal\left((\delta s_{n})^4\right)} ,
\end{align}
where we defined
\begin{align}
	\hat{\chi} &= f(\bar{s}_{n})g(\bar{s}_{n})\frac{ (-i \delta s_{n} \tau)^2}{2 \hbar^2} [\hat{H}_{\mathrm{fin} } ,\hat{H}_{\mathrm{ini}}] + 
	f(\bar{s}_{n})g^2(\bar{s}_{n})\frac{ (-i \delta s_{n} \tau)^3}{12 \hbar^3} \big[\hat{H}_{\mathrm{fin} } ,[\hat{H}_{\mathrm{fin} } ,\hat{H}_{\mathrm{ini}}]\big]
-
	f^2(\bar{s}_{n})g(\bar{s}_{n})\frac{ (-i \delta s_{n} \tau)^3}{12 \hbar^3} \big[\hat{H}_{\mathrm{ini} } ,[\hat{H}_{\mathrm{fin} } ,\hat{H}_{\mathrm{ini}}]\big] 
	\nonumber \\
	&= f(\bar{s}_{n})g(\bar{s}_{n}) \left[\frac{ (-i \delta s_{n} \tau)^2}{2 \hbar^2} \hat{A} + 
	\frac{ (-i \delta s_{n} \tau)^3}{12 \hbar^3}\hat{B}(\bar{s}_{n})\right] ,
\end{align}
with $\hat{A} = [\hat{H}_{\mathrm{fin} } ,\hat{H}_{\mathrm{ini}}]$ and $\hat{B}(\bar{s}_{n}) =\big[g(\bar{s}_{n})\hat{H}_{\mathrm{fin} }- f(\bar{s}_{n})\hat{H}_{\mathrm{ini} } ,[\hat{H}_{\mathrm{fin} } ,\hat{H}_{\mathrm{ini}}]\big]$. Now, we observe that if the condition 
\begin{align}
	\norm{\hat{\chi}} \ll \tau \delta s_{n}\norm{\hat{H}(\bar{s}_{n})}/\hbar \label{Ap:Eq:Inequality1}
\end{align}
is satisfied, then we can write
\begin{align}
	e^{- \frac{i\tau \delta s_{n}}{\hbar}\hat{H}(\bar{s}_{n})} \approx e^{- \frac{i\tau \delta s_{n}}{\hbar}g(\bar{s}_{n}) \hat{H}_{\mathrm{fin}} }e^{- \frac{i\tau \delta s_{n}}{\hbar}f(\bar{s}_{n}) \hat{H}_{\mathrm{ini}} } 
	,
\end{align}
up to a correction factor  $\Ocal\big((\delta s_{n})^4\big)$. In this way, we can use the triangle inequality to write
\begin{align}
	\norm{\hat{\chi}} = \abs{g(\bar{s}_{n})f(\bar{s}_{n})} \left\vert\left\vert\frac{(i\tau \delta s_{n})^2}{2 \hbar^2} \hat{A} + \frac{(i\tau \delta s_{n})^3}{12 \hbar^3}\hat{B}(\bar{s}_{n}) \right\vert\right\vert 
	\leq \abs{g(\bar{s}_{n})f(\bar{s}_{n})}\frac{(\tau \delta s_{n})^2}{2 \hbar^2}\left(\norm{ \hat{A} } + \frac{\tau \delta s_{n}}{6 \hbar}\norm{\hat{B}(\bar{s}_{n})}\right) ,
\end{align}
which can be used in Eq.~\eqref{Ap:Eq:Inequality1} to give us the final inequality
\begin{align}
	\frac{\tau \delta s_{n}}{\hbar}\norm{\hat{H}(\bar{s}_{n})} \gg  \abs{g(\bar{s}_{n})f(\bar{s}_{n})}\frac{(\tau \delta s_{n})^2}{2 \hbar^2}\left(\norm{ \hat{A} } + \frac{\tau \delta s_{n}}{6 \hbar}\norm{\hat{B}(\bar{s}_{n})}\right). \label{Ap:Eq:Inequality}
\end{align}

The above inequality can be solved through the solution of the generic equation given by $b x^3 + a x^2 - c x \ll 0$, in which the quantities $a,b,c$ and the variable $x$ are positive definite numbers. The admissible solution for this problem is obtained as $0 < x \ll (-a + \sqrt{a^2 + 4bc})/2b$, and therefore it allows us to write the admissible solution for $\delta s_{n}$ of the Eq.~\eqref{Ap:Eq:Inequality} as
\begin{align}
	\tau \delta s_{n} \ll \frac{\hbar  \sqrt{f(\bar{s}_{n})g(\bar{s}_{n}) \left(9 \norm{ \hat{A} }^2 f(\bar{s}_{n})g(\bar{s}_{n})+12 \norm{\hat{B}(\bar{s}_{n})} \norm{\hat{H}(\bar{s}_{n})}\right)}}{\norm{\hat{B}(\bar{s}_{n})} f(\bar{s}_{n})g(\bar{s}_{n})}
	- \frac{3 \norm{ \hat{A} } \hbar }{\norm{\hat{B}(\bar{s}_{n})}} \label{ApEq:reference}.
\end{align}

It is worth mentioning here that the above equation can be rewritten in a compact way as $\delta s_{n} = \tau_{0} / \tau$, with $\tau_{0}$ an digitized adiabatic reference time satisfying
\begin{align}
	\tau_{0} \ll \frac{\hbar  \sqrt{f(\bar{s}_{n})g(\bar{s}_{n}) \left(9 \norm{ \hat{A} }^2 f(\bar{s}_{n})g(\bar{s}_{n})+12 \norm{\hat{B}(\bar{s}_{n})} \norm{\hat{H}(\bar{s}_{n})}\right)}}{\norm{\hat{B}(\bar{s}_{n})} f(\bar{s}_{n})g(\bar{s}_{n})}
	- \frac{3 \norm{ \hat{A} } \hbar }{\norm{\hat{B}(\bar{s}_{n})}} \label{ApEq:reference2} .
\end{align}

Therefore, following the strategy of obtaining a general and robust condition for the adiabatic regime imitation, we first assume the continuum of values for the above equation, replacing $\bar{s}_{n} \rightarrow s$, and we also consider an estimate of a reasonable total run time by invoking the condition for the adiabatic theorem to write (
without loss of generality we use the inequality $\tau \geq \tau_{\mathrm{ad}}$, but the adiabatic condition $\tau \gg \tau_{\mathrm{ad}}$ also can be used here to get the same final result
)
\begin{align}
	\tau \geq \tau_{\mathrm{ad}} = \max_{0\leq s \leq 1} \hbar \frac{\norm{d_{s}H(s)}}{\Delta^2(s)}  .
\end{align}

Through these two considerations, we find a general condition on each time interval $\delta s_{n}$ for digitized adiabatic protocol as
\begin{align}
	\delta s_{n} \ll \delta s_{\mathrm{ad}} = \frac{1}{\tau_{\mathrm{ad}}}\min_{s} \left(\frac{\sqrt{f(s)g(s)\eta(s) }}{\norm{\hat{B}(s)} f(s)g(s)}
	-\frac{3 \norm{ \hat{A} } }{\norm{\hat{B}(s)}}\right) .
\end{align}
where $\eta(s) = 9 \norm{ \hat{A} }^2 f(s)g(s)+12 \norm{\hat{B}(s)} \norm{\hat{H}(s)}$.

\twocolumngrid


%

\end{document}